\documentclass{article}

\begin{document}

\title{Understanding weak values without new probability theory}
\author{Riuji Mochizuki\thanks{E-mail:rjmochi@tdc.ac.jp\ \ telephon:+81-3-6380-9308}\\
Laboratory of Physics, Tokyo Dental College,\\ 2-9-7 Kandasurugadai, Chiyoda-ku, Tokyo 101-0062, Japan }

\maketitle

\begin{abstract}
The physical meaning of weak values and measurements can be completely understood with Born rule and the general probability theory. It is known that the weak value of an observable $\hat A$ with post-selection $\langle F|$ may be out of the eigenvalue range of $\hat A$. This is because the weak value of $\hat A$ with the post-selection is, in general, not the expectation value of $\hat A$, but the expectation value of $\hat A| F\rangle\langle F|$ boosted by the post-selection.\\
{\bf keywords}\ weak value, weak measurement, post-selection, anomalous weak value, Born rule
\end{abstract}

Nearly three decades have passed since Aharonov et al. \cite{Aha} introduced weak measurements and values.  Nevertheless, they remain a subject of debate.  Recently, Vaidman \cite{Vaid1, Vaid2} analyzed the nested Mach-Zehnder interferometer experiment with two-state vector formalism and insisted that the past of a quantum particle could be described according to the weak trace.  Li et al. \cite{Li, Salih} challenged Vaidman's claim and insisted that the weak trace could be understood without any unusual probability theory if the disturbances of the weak measurements are considered.  However, they agreed with Vaidman with regard to the physical meaning of the weak values.   

Moreover, Ferrie and Combes \cite{Ferrie1, Ferrie2} argued that weak values are classical statistic quantities, which gave rise to a number of rebuttals [8 - 12].  In particular, Pusey \cite{Pusey} showed that anomalous (imaginary, negative, and unbounded) weak values are non-classical and proofs of contextuality.  However, he did not show how the contextuality is responsible for the anomalous weak values.

As confirmed by many experiments, the measured value of the weak measurement agrees with the corresponding weak value.  In this paper, therefore, we carefully examine the process of the weak measurement to know what the weak value is.  It is shown that the physical meanings of weak measurements and weak values can be completely understood within the framework of a conventional quantum mechanical approach, that is, with Born rule and the general probability theory.  Much confusion concerning the weak value has been caused by the following hypothesis: the weak value of $\hat A$ is a conditional or some kind of expectation value of $\hat A$.  We demonstrate 
\begin{equation}
\langle \hat A\rangle^w\equiv\frac{\langle F|\hat A|I\rangle}{\langle F|I\rangle}\label{eq:wv}
\end{equation}
is not the expectation value of $\hat A$ with the pre-state $|I\rangle$ and post-state $\langle F|$; its real and imaginary parts are, which are accompanied with some constant factors, essentially the expectation values of $(1/2)(|F\rangle\langle F|\hat A+\hat A|F\rangle\langle F|)$ and $(i/2)(|F\rangle\langle F|\hat A-\hat A|F\rangle\langle F|)$ for $|I\rangle$ boosted by $1/\big |\langle I|F\rangle \big|^2$ via the post-selection, respectively.   If $\hat A$ and $|F\rangle\langle F|$ do not commute, these values are completely different from the real and imaginary parts of the expectation value of $\hat A$ for $|I\rangle$ with the post-selection.  Moreover, even if $\hat A$ is a projection operator, $\hat A|F\rangle\langle F|$ is not.  Therefore, we have no reason to expect the weak value of $\hat A$ within its eigenvalue range.

First, we examine the process of the weak measurement by means of von Neumann-type measurement \cite{von} according to \cite{Aha}.  The interaction Hamiltonian $\hat H_A$ between an observable $\hat A$ of the observed system and the momentum $\hat \pi_A$ of the pointer of the measuring device is  
\begin{equation}
\hat H_A\equiv g_A\hat A\hat \pi_A,\label{eq:Hamiltonian}
\end{equation}
where $g_A$ is the coupling constant.   $\hat H_A$ is assumed to be constant and roughly equivalent to the total Hamiltonian $\hat H$ over some interaction time $t_A$.   The initial wavefunction $\phi_A (x)$ of the measuring apparatus is assumed to be
\begin{equation}
\phi_A (x)=\langle x_A|\phi_A\rangle =\Big({1\over \sqrt{2\pi}\sigma_A}\Big)^{1/2}\exp\Big( -{x_A^2\over 4\sigma_A^2}\Big),
\label{eq:wave}
\end{equation}
where $x_A$ is the position of the pointer of the measuring device.  The initial state $|\Phi_A(0)\rangle=|I\rangle |\phi_A\rangle$, where $|I\rangle$ is the initial state of the observed system, of the unified system of the observed system and the measuring device, evolves unitarily obeying the Schr\"{o}dinger equation:
\begin{equation}
i\hbar{d\over dt}|\Phi_A(t_A)\rangle =\hat H|\Phi_A (t_A)\rangle \sim\hat H_A|\Phi_A (t_A)\rangle ,
\end{equation}
and becomes
\begin{equation}
|\Phi_A(t_A)\rangle =\exp\Big(-\frac{i\hat H_At_A}{\hbar}\Big)|\Phi_A(0)\rangle.\label{eq:full}
\end{equation}
Up to the first order of $g_At_A$,  
\begin{equation}
|\Phi_A (t_A)\rangle =|I\rangle |\phi_A\rangle -{ig_At_A\over\hbar}\hat A|I\rangle\hat\pi_A |\phi_A\rangle  .\label{eq:gtichiji}
\end{equation}
Instead, we can equally describe the unified system by means of the density matrix
\begin{equation}
\hat\rho_A(t_A)=|\Phi_A(t_A)\rangle\langle\Phi_A(t_A)|.\label{eq:matrix}
\end{equation}
Without any post-selection, the expectation value of $\overline x_A$ of the pointer's position $\hat x_A$ for this state is
\begin{equation}
\begin{array}{rl}
\overline{x}_A&={\rm Tr}\big [\hat \rho_A(t_A)\hat x_A\big ]\\
&=g_At_A\langle I|\hat A|I\rangle .
\end{array}
\end{equation}

In \cite{Aha}, it was insisted that the state of the measuring device right after the unitary interaction with the measured system and with post-selection $\langle F|$ for the measured system  is 
\begin{equation}
\frac{\langle F|\Phi_A(t)\rangle}{\langle F|I\rangle}=|\phi_A\rangle -\frac{ig_At_A}{\hbar}\frac{\langle F|\hat A|I\rangle}{\langle F|I\rangle}\hat \pi_A|\phi_A\rangle.\label{eq:shuchou}
\end{equation}
Here, we show that this claim is not exact because of the non-separability of the measured system and the measuring device \cite{Mochi, Esp}.  To this end, we assume that the ensemble $S$ of the observed system and the ensemble $M$ of the measuring device after their unitary interaction are both separately obtained by combining all the elements of sub-ensembles, each of which is described by its own ket.  Then, each element of $S$ belongs to one of the sub-ensembles $E_{i},\ i =1,2,\cdots$ described by $|s_{i}\rangle$ and each element of $M$ belongs to one of the  sub-ensembles $E_{\alpha},\ {\alpha}=1,2,\cdots$ described by $|m_{\alpha}\rangle$, such that the sub-ensemble $\varepsilon_{i,\alpha}$ of the unified system, whose elements belong to both $E_i$ and $E_{\alpha}$, is described by the density matrix 
\begin{equation}
\hat\rho_{i,\alpha}=|s_{i}\rangle |m_{\alpha}\rangle\langle m_{\alpha}|\langle s_{i}|.
\end{equation}
Because the unified system's ensemble $\varepsilon$ is the union of all the $\varepsilon_{i,\alpha}$, the density matrix $\hat\rho^{\prime}$ describing $\varepsilon$ should be written as the weighted sum of all the $\hat\rho_{i,\alpha}$:
\begin{equation}
\hat\rho^{\prime}=\sum_{i,\alpha}P_{i,\alpha}\hat\rho_{i,\alpha },
\end{equation}
where $P_{i,\alpha}$ are suitable factors.  However, $\varepsilon$ is defined to be described by (\ref{eq:gtichiji}), such that it should be described by the density matrix (\ref{eq:matrix}).  $\hat\rho_A(t)$ and $\hat\rho^{\prime}$ are necessarily different, except in the case that $|\Phi_A(t)\rangle$ is a product of a vector $|S\rangle$ in the Hilbert space of  the observed system and a vector $|M\rangle$ in the Hilbert space of the measuring apparatus, i.e.,
\begin{equation}
|\Phi_A(t)\rangle =|S\rangle |M\rangle.
\end{equation}
(\ref{eq:gtichiji}) does not have this form.  Therefore, the previous assumption has been shown to be false.

We must say for the above reason that both the observed system and the measuring device do not have separate ensembles of their own.  Therefore, we conclude that the operation of $\langle F|$ on (\ref{eq:gtichiji}) changes the unified system and (\ref{eq:shuchou}) is not the state of the measuring device right after their unitary interaction, i.e. right after $t_A$.

Then, we clarify what the weak value is.  This requires careful examination of the weak measurement, especially of the post-selection.  With this end in view we must consider two measuring devices: one weakly measures the observable $\hat A$ and the other selects the post-state $\langle F|$ via a projection measurement.  Their interaction Hamiltonians are (\ref{eq:Hamiltonian}) and
\[
\hat H_F=g_F\hat F\hat\pi_F,
\]
where $\hat F\equiv |F\rangle\langle F|$ and $\hat\pi_{F}$ is the momentum of the pointer of the measuring device of $\hat F$. The initial state of the unified system of the observed system and the two measuring devices is
\[
|\Phi(0)\rangle=|I\rangle |\phi_A\rangle |\phi_F\rangle,
\]
where $|\phi_F\rangle$ is the initial state of the measuring device of $\hat F$ whose wave function is assumed to be
\begin{equation}
\phi_{F} (x)=\langle x_F|\phi_{F}\rangle =\Big({1\over \sqrt{2\pi}\sigma_{F}}\Big)^{1/2}\exp\Big( -{x_{F}^2\over 4\sigma_{F}^2}\Big),
\label{eq:waveA}
\end{equation}
where $x_F$ is the position of the pointer of the measuring device of $\hat F$.

We weakly measure $\hat A$ and then select the final state.  Therefore, the state following the interaction between the observed system and the measuring devices is
\begin{equation}
|\Phi(t)\rangle =\exp \Big(-\frac{iH_Ft_F}{\hbar}\Big)\exp \Big(-\frac{iH_At_A}{\hbar}\Big)|\Phi(0)\rangle,
\end{equation}
where
\[
t=t_A+t_F.
\]
Up to the first order of $g_At_A$,
\begin{equation}
|\Phi(t)\rangle =\exp\Big(-\frac{i\hat H_Ft_F}{\hbar}\Big)|\phi_F\rangle
\Big[|I\rangle |\phi_A\rangle -{ig_At_A\over\hbar}\hat A|I\rangle\hat\pi_A |\phi_A\rangle\Big]  .\label{eq:gtichijinoni}
\end{equation}

We define the partial density matrix $\hat \rho^{(m)}(t)$ of the measuring devices as
\begin{equation}
\hat \rho^{(m)}(t)={\rm Tr}^{(s)}\big[|\Phi(t)\rangle\langle\Phi(t)|\big],\label{eq:density}
\end{equation}
where ${\rm Tr}^{(s)}$ is the partial trace of the observed system.  By calculating the expectation value of either $\hat x_A$ or $\hat x_F$, we can obtain the expectation value of either $\hat A$ or $\hat F$ accurately as follows: 
\begin{equation}
\overline x_A\equiv {\rm Tr}\big[\hat\rho^{(m)}(t)\hat x_A\big]
=g_At_A\langle I|\hat A|I\rangle,
\end{equation}
\begin{equation}
\overline{x}_F\equiv {\rm Tr}\big[\hat\rho^{(m)}(t)\hat x_F\big]=g_Ft_F\langle I|\hat F|I\rangle.\label{eq:Fdake}
\end{equation}

Because $\hat x_A$ and $\hat x_F$ commute, we can obtain their measured values $X_A$ and $X_F$ simultaneously.  However, we cannot know the expectation values of both $\hat A$ and $\hat F$ simultaneously\cite{Arthurs}.  Its reason is almost the same as the previous discussion: If the ensembles $M_A$ and $M_F$ of the two measuring devices after their unitary interaction with the measured system are both separately obtained by combining all the elements of the sub-ensembles, each of them can be described by its own ket.  Each element of $M_A$ belongs to one of the sub-ensembles $E_{\alpha},\ \alpha =1,2,\cdots$, described by $|a_{\alpha}\rangle$ and each element of $M_F$ belongs to one of the sub-ensembles $E_{\beta},\ \beta =1,2,\cdots$, described by $|f_{\beta}\rangle$ such that the sub-ensemble $\varepsilon_{\alpha,\beta}$ of the combined measuring device, whose elements belong to both $E_{\alpha}$ and $E_{\beta}$, is described by the density matrix
\[
\hat\rho_{\alpha ,\beta}=|f_{\beta}\rangle |a_{\alpha}\rangle\langle a_{\alpha}|\langle f_{\beta}|,
\]
and the ensemble of the combined measuring device is described as the weighted sum of $\hat\rho_{\alpha,\beta}$:
\begin{equation}
\hat\rho^{\prime\prime}=\sum_{\alpha ,\beta}P_{\alpha ,\beta}\hat\rho_{\alpha ,\beta},\label{eq:simulrhoprime}
\end{equation}
where $P_{\alpha ,\beta}$ are suitable factors.  However, (\ref{eq:density}) does not take the form of (\ref{eq:simulrhoprime}) if $\hat F$ and $\hat A$ do not commute.  Therefore, $\hat x_A$ and $\hat x_F$ are entangled, i.e., the position operators of both measuring devices after the unitary interaction with the measured system do not have their own separate ensembles.   We should regard the measurement of $\hat x_A$ and $\hat x_F$ as {\it one} manipulation.

Then, we reconsider the process to know what outcome we obtain, i.e., what observable of the {\it unified} measuring device we read in this manipulation and what observable of the observed system corresponds to the outcome of the unified measuring device.

 Although both $\hat x_A$ and $\hat x_F$ are measured in the weak measurement with post-selection, their measured values $X_A$ and $X_F$ should not be treated separately, as shown above.  Because $\hat x_F$ is a projection operator, $X_F$ is $1$ or $0$ and $X_F^n=X_F\ (n\ne0)$.  Here and hereafter, we put $g_Ft_F=1$. On the other hand, we can know only the {\it sum} of post-selected (and {\it not} selected) $X_A$'s, so that the outcome must be regarded as linear of $X_A$.  Therefore the outcome of the weak measurement with the post-selection is $X_AX_F$ and the measured observable is $\hat x_A\hat x_F$.  Its expectation value is
\begin{equation}
\begin{array}{rl}
\overline{x_Ax_F}&={\rm Tr}\big[ \hat x_F\hat x_A\hat\rho^{(m)}(t)\big]\\
&={\rm Tr}\big[ \hat x_A\hat x_F\hat\rho^{(m)}(t)\big]\\
&=\frac{1}{2}g_At_A\langle I|(\hat F\hat A+\hat A\hat F)|I\rangle,\label{eq:FAAF}
\end{array}
\end{equation}
which is equal to $\langle X_AX_F\rangle$, the average of $X_AX_F$.  Because $\hat x_F$ and $\hat x_A$ are entangled and $\overline{x_Ax_F}\ne\overline x_A\cdot\overline x_F$, we cannot obtain the expectation value of $\hat A$ if it does not commute with $\hat F$.  (We can approximately obtain the expectation value of $\hat F$ because the first measurement is weak.)  Instead, we can obtain the expectation value of  $(1/2)(\hat F\hat A+\hat A\hat F)$ via the weak measurement.  

 The physical meaning of post-selection should be considered carefully in this context.  In the post-selection, we select cases of $X_F=1$, which is approximate selection of the final state $\langle F|$.  Because  the post-selection $X_F=1$ (i.e., $X_F\ne 0$) implies $X_AX_F\ne 0$ (if $X_A\ne 0$),  
 the average of $X_AX_F$ after the post-selection $X_F=1$ is equal to the average of $X_A$ after the post-selection:
\begin{equation}
\langle X_A\rangle^{(p)}=\langle X_AX_F\rangle^{(p)},
\end{equation}
where $\langle\ \rangle^{(p)}$ stands for the average after post-selection.  Moreover, because $\langle X_AX_F\rangle^{(p)}$ is the quotient of the sum of post-selected $X_AX_F$'s, which is equal to the sum of all $X_AX_F$'s without any post-selection,  divided by the number of the post-selected data, it is boosted by $1/\langle X_F\rangle$:
\begin{equation}
\frac{\langle X_AX_F\rangle^{(p)}}{\langle X_AX_F\rangle}=\frac{1}{\langle X_F\rangle},
\end{equation}
where $\langle X_F\rangle$ is nearly equal to $\overline x_F$, because the first measurement is weak.  For example, if the measured values are 
\begin{equation}\begin{tabular}{|c|c|c|c|c|c|} \hline $X_A$ & 2 & 2 & 2 & 2 & 2 \\ \hline $X_F$ & 1 & 0 & 0 & 0 & 0 \\ \hline
\end{tabular},
\end{equation}
then, $\langle X_AX_F\rangle =0.4$, $\langle X_F\rangle =0.2$, $\langle X_AX_F\rangle^{(p)}=2$. 

Gathering these pieces, we obtain 
\begin{equation}
\langle X_A\rangle^{(p)}=\frac{\overline{x_Ax_F}}{\overline x_F}.\label{eq:average}
\end{equation}
By means of (\ref{eq:Fdake}) and (\ref{eq:FAAF}), (\ref{eq:average}) becomes
\begin{equation}
\frac{\langle X_A\rangle^{(p)}}{g_At_A}=\frac{\langle I|(\hat F\hat A+\hat A\hat F)|I\rangle}{2\langle I|\hat F|I\rangle}.\label{eq:real}
\end{equation}
The right-hand side of (\ref{eq:real}) is the real part of the weak value (\ref{eq:wv}). If some pairs of $\hat A$, $\hat F$ and $|I\rangle\langle I|$ commute, it becomes $\langle I|\hat A|I\rangle$ independently of the post-selection.  If $\big[\hat A,[\hat F,|I\rangle\langle I|]\big]=0$, it is in proportion to $\langle I|\hat A|I\rangle$. Otherwise, it is not an expectation value of $\hat A$ in any sense, less to be the expectation value of $\hat A$ after the post-selection $\langle F|$.  In contrast, it is the expectation value of $(1/2)(\hat F\hat A+\hat A\hat F)$ boosted by the post-selection.  This is the reason why the weak value of $\hat A$ may be out of the eigenvalue range of $\hat A$.

In summary, our main result comes down to (\ref{eq:average}), which clarifies that weak values can be completely understood within the framework of conventional quantum mechanics, that is, with Born rule and the general probability theory.  Weak measurement with post-selection should be considered as a method to measure an observable which are product of two observables, one of which is a projection operator, and to boost its measured value.

\end{document}